# A class of 5D inhomogeneous models with a cosmological constant[†]

**Pantelis S. Apostolopoulos**[1]

[1] Department of Environment, Ionian University, Mathematical Physics and Computational Statistics Research Laboratory, Zakynthos 29100, Greece

**Abstract:** In this work we would like to address the problem of the effect of bulk matter on the brane cosmological evolution in a general way. We assume that the spatial part of the brane metric is not maximally symmetric, therefore spatially inhomogeneous. However we retain the conformal flatness property of the standard cosmological model (FRW) i.e. the Weyl tensor of the induced 4D geometry is zero. We refer to it as Spatially Inhomogeneous Irrotational (SII) brane. It is shown that the model can be regarded as the 5D generalization of the SII spacetimes found recently [1].

## 1. Introduction

In the context of the Randall-Sundrum model [2], the Universe is identified with a four-dimensional hypersurface (a 3-brane) in a five-dimensional bulk with negative cosmological constant (AdS space). The pertinent action is:

$$S = \int d^5 x \sqrt{-\hat{g}} \left( \Lambda + M^3 \hat{R} + \mathbf{L}_B \right) + \int d^4 x \sqrt{-g} \left( -V + \mathbf{L}_b \right), \tag{1}$$

where $\hat{R}$ is the curvature scalar of the five-dimensional bulk metric $\hat{g}_{AB}$, $-\Lambda$ the bulk cosmological constant ($\Lambda>0$), $V$ the brane tension, and $g_{\alpha\beta}$ the induced metric on the brane (we neglect higher curvature invariants in the bulk and induced gravity terms on the brane). The Lagrangian density $\mathbf{L}_B$ describes the matter content (particles or fields) of the bulk, while the density $\mathbf{L}_b$ describes matter localized on the brane.

The possibility that our Universe is identified with a 4D hypersurface (3-brane) embedded in a higher-dimensional bulk space allows for interesting novel features in the cosmological evolution as perceived by a brane observer. In addition, the 3-brane must also have a tension that balances the effect of the negative cosmological constant on the expansion. For low energy densities, the effective Friedmann equation on the brane has the standard form. Several features and variations of this scenario have been considered. (For recent reviews see [3].)

The geometry is non-trivial (warped) along the fourth spatial dimension, so that an effective localization of low-energy gravity takes place near the brane. (No such localization takes place for the bulk matter.) For low matter densities on the brane and a pure AdS bulk (no bulk matter), the cosmological evolution as seen by a brane observer reduces to the standard Friedmann-Robertson-Walker cosmology. Therefore the brane is assumed to be normal to the unit spacelike vector field $w^A$ that is tangential to the extra spatial dimension. In the standard brane cosmology, the metric is parametrized as

$$ds^2 = -m^2(\tau, w)d\tau^2 + a^2(\tau, w)d\Omega_k^2 + dw^2, \qquad (2)$$

with $m(\tau, w=0) = 1$.
The brane is located at $w=0$, while we identify the half-space $w>0$ with the half-space $w<0$.

We are interested in a more general background where the brane is conformally flat but not Friedmann-Robertson-Walker hence the geometry is inhomogeneous assuming the metric:

$$ds^2 = dw^2 + A^2 dx^2 + B^2\left(-dt^2 + dy^2 + dz^2\right) \qquad (3)$$

where $A(t,w,x,y,z), B(t,w,x,y,z)$ are functions of their arguments.
The bulk equations of motion have the form (throughout this work we use the following index conventions: bulk 5D indices are denoted by capital Latin letters $A, B, \ldots = 0, 1, 2, 3, 4$ and Greek letters denote space-time indices $\alpha, \beta, \ldots = 0, 1, 2, 3$):

$$G_B^A = \frac{1}{2M^3}\left(T_B^A + \Lambda \delta_B^A\right), \qquad (4)$$

where $T_B^A$ denotes the total energy-momentum tensor, i.e.

$$T_{AB} = T_{AB}^{\text{BULK}} + \delta(w)\left(T_{AB}^{\text{BRANE}} + V g_{AB}\right) \qquad (5)$$

The modified 4D field equations are derived by assuming a $Z_2$ symmetry of the bulk around the brane and employing Israel's junction conditions [3]:

$$G_{\alpha\beta} = -3\lambda g_{\alpha\beta} + \frac{V}{24M^6}T_{\alpha\beta} + \frac{1}{4M^6}\mathbf{S}_{\alpha\beta} - \mathbf{E}_{\alpha\beta} + \frac{1}{3M^3}\mathbf{F}_{\alpha\beta},$$

where

$$\mathbf{S}_{\alpha\beta} = \frac{1}{2}TT_{\alpha\beta} - \frac{1}{4}T_{\alpha\gamma}T_\beta^\gamma + \frac{3T_{\gamma\delta}T^{\gamma\delta} - T^2}{24}g_{\alpha\beta}$$

$$\mathbf{F}_{\alpha\beta} = T_{AB}^{\text{BULK}} g_\alpha^A g_\beta^B + \left(T_{AB}^{\text{BULK}} w^A w^B - \frac{T^{\text{BULK}}}{4}\right)g_{\alpha\beta}$$

$$\mathbf{E}_{\alpha\beta} = \mathbf{E}_{AB} g_\alpha^A g_\beta^B = C_{ACBD} w^C w^D g_\alpha^A g_\beta^B$$

and $\lambda = (V^2/12M^3 - \Lambda)/12M^3$ is the effective cosmological constant.

We observe that, apart from the terms quadradic to the brane energy-momentum tensor, there exist two additional terms corresponding to: a) the projection of the 5D Weyl tensor $\mathbf{E}_{\alpha\beta}$ and b) the projected (normal to $w^A$) tensor $\mathbf{F}_{\alpha\beta}$ that contains the bulk matter contribution. Since both tensors are 5D objects we conclude that *both induce bulk* *effects on the brane*. In the case of an empty bulk, the 5D contributions on the brane are coming from the non-local effects of the free gravitational field incorporated in $\mathbf{E}_{\alpha\beta}$ (5D bulk gravitons).

For the rest of the paper we use geometrized units such that $8\pi G = c = 1$ and the Planck scale equal to unity.

## 2. Methods

As the generality of the underlying geometry is increased, equations (4) and the resulting equation of motion become progressively highly non-linear and often lead to models without clear physical meaning. Therefore the inspection of simplification properties allows us to examine the possibility of the existence of exact solutions with sound physical interest. In addition, the coupling of the geometry with the dynamics, through the EFEs, implies that there is a mutual influence between any geometrical and dynamical constraint with subsequent restrictions in the structure of the corresponding models.

The key kinematical and dynamical aspects that we assume in the present setup can be collected as follows (see e.g. [1] for a 4-dimentional perspective):

1. The spacelike vector field $w^A$ is geodesic and vorticity-free i.e. $\dot{w}_A = w_{A;B} w^B = 0 = w_{[A;B]} = \nabla_{[B} w_{A]}$.
2. The symmetric tensor $\hat{h}_{AB} = \hat{g}_{AB} - w_A w_B$ is the proper metric of the distribution $\mathbf{w} = \mathbf{0}$ with a well-defined covariant derivative $D_A$ that satisfies $D_C \hat{h}_{AB} = \hat{h}_C^K \hat{h}_A^I \hat{h}_B^J \nabla_K \hat{h}_{IJ} = 0$.
3. There exist 4 independent unit vector fields $\{\mathbf{u}, \mathbf{x}, \mathbf{y}, \mathbf{z}\}$ that are hypersurface orthogonal $u_{[A} u_{B;C]} = x_{[A} x_{B;C]} = y_{[A} y_{B;C]} = z_{[A} z_{B;C]} = 0$. For *each pair* e.g. $\{\mathbf{w}, \mathbf{x}\}$ the second order symmetric tensor $\hat{p}_{AB} = \hat{h}_{AB} - x_A x_B$ represents the proper metric of the screen space $\mathbf{S}$, which is defined as $\mathbf{w} \wedge \mathbf{x} = \mathbf{0}$, with associated 3D derivative $\overline{\nabla}_A$. In particular $\overline{\nabla}_C p_{AB} = p_C^K p_A^I p_B^J \nabla_K p_{IJ} = 0$ and $\{w^A, x^A\}$ are *surface forming* i.e. $p_A^K L_\mathbf{w} x_K = 0$.
4. The Weyl tensor of the hypersurfaces *w*=const. vanishes.

As a result one should check for the existence of Intrinsic Conformal Vector Fields (ICVFs) that generate transformations preserving the conformal structure of the 3D screen space $\mathbf{S}$ i.e. vector fields $\mathbf{X}$ satisfying:

$$p_A^C p_B^D \mathbf{L}_\mathbf{X} p_{CD} = \overline{\nabla}_{(B} X_{A)} = 2\phi(\mathbf{X}) p_{AB} \tag{6}$$

where $p_A^K X_K = X_A$ i.e. $\mathbf{X}$ are lying in the 3D screen space $\mathbf{S}$.

Assuming that the bulk energy momentum tensor has the form:

$$T_{AB}^{BULK} = p_\| w_A w_B \tag{7}$$

the general solution of (4) is:

$$ds^2 = dw^2 + \left\{\frac{D[\ln(D/E)]_{,x}}{\sqrt{\varepsilon + F(x)}}\right\}^2 dx^2 + \frac{D^2}{E^2}\left(-dt^2 + dy^2 + dz^2\right) \tag{8}$$

Here $D(w,x)$, $Y(x)$, $Z(x)$, $T(x)$, $C(x)$ and $F(x)$ are functions of their arguments, $\varepsilon = \pm 1$ ($\neq 0$) corresponds to the constant curvature of the hypersurfaces $w, x = $ const. and we have set:

$$E(t,x,y,z) = C \cdot \left\{1 + \frac{\varepsilon}{4C^2}\left[(y-Y)^2 + (z-Z)^2 - (t-T)^2\right]\right\}. \tag{9}$$

Defining the function $k = \varepsilon/C^2$, we obtain

$$E(t,x,y,z) = C \cdot \left\{1 + \frac{k}{4}\left[(y-Y)^2 + (z-Z)^2 - (t-T)^2\right]\right\} \tag{10}$$

and the metric becomes:

$$ds^2 = dw^2 + \left\{\frac{D[\ln(D/E)]_{,x}}{\sqrt{\varepsilon + F(x)}}\right\}^2 dx^2 + \frac{D^2}{E^2}\left(-dt^2 + dy^2 + dz^2\right) \tag{11}$$

where a "," denotes partial differentiation w.r.t. to the following coordinate.

The metric (8) is the general solution of (4) provided that the following equation holds:

$$F(x) + DD_{,ww} + (D_{,w})^2 - \frac{\Lambda}{3}D^2 = 0. \tag{12}$$

In addition the energy-momentum conservation equation:

$$G^A_{B;A} = 0 \Leftrightarrow (p_{\|})_{;A} w^A + w^A_{;A} = 0 \tag{13}$$

is satisfied identically as we can easily check computationally.

## 3. Results and Discussion

The metric (8) or (11) manifests the quasi-symmetry of the 5D bulk spacetime. In fact, it can be shown that (11) admits a *10-dimensional* Lie algebra of ICVFs consisting of 6 independent intrinsic Killing Vector Fields and 4 intrinsic *gradient proper* Conformal Vector Fields. However, in general, *does not admit* any global symmetry. The brane ($w = 0$) is conformally flat (but not FRW) and can be seen similar to Stephani's universes. The $p_{\|}$ pressure can be assigned to any moduli fields in the bulk and live only in the extra dimension and the brane-bulk energy exchange is attributed to those fields [4]. The brane metric is, in general, SII but still can be asymptotically FRW. Due to the inhomogeneity of the brane metric, the fluid could (must) be anisotropic and viscous, therefore give us the opportunity to check various possibilities on the effect of inhomogeneities coming from the extra dimension.

**Funding:** This research received no external funding.
**Institutional Review Board Statement:** Not applicable.
**Informed Consent Statement:** Not applicable.
**Data Availability Statement:** Not applicable here.